\documentclass[10pt,letterpaper]{article}
\usepackage[top=0.85in,left=2.75in,footskip=0.75in]{geometry}

\usepackage{amsmath,amssymb}
\usepackage{tabularx} 
\usepackage{siunitx} 
\DeclareSIUnit\Molar{\textsc{m}} 
\newcommand{\prob}{\mathbb{P}}
\newcommand{\ca}{Ca$^{2+}$}

\newcommand{\ipthree}{IP$_3$}
\newcommand{\ipr}{\ipthree R}

\newcommand{\Jleak}{J_{\text{leak}}}
\newcommand{\Jr}{J_{r}}
\newcommand{\kOn}{k_{\text{on}}}
\newcommand{\kOff}{k_{\text{off}}}

\usepackage{changepage}
\usepackage{todonotes}

\usepackage{textcomp,marvosym}

\usepackage{cite}

\usepackage{nameref,hyperref}

\usepackage[right]{lineno}

\usepackage[nopatch=eqnum]{microtype}
\DisableLigatures[f]{encoding = *, family = * }

\usepackage{array}

\usepackage{subfig}

\newcolumntype{+}{!{\vrule width 2pt}}

\newlength\savedwidth

\usepackage{setspace} 
\raggedright
\usepackage[aboveskip=1pt,labelfont=bf,labelsep=period,justification=raggedright,singlelinecheck=off]{caption}

\makeatletter
\renewcommand{\@biblabel}[1]{\quad#1.}
\makeatother

\usepackage{lastpage,fancyhdr,graphicx}
\usepackage{epstopdf}
\fancyheadoffset[L]{2.25in}
\fancyfootoffset[L]{2.25in}
\newcommand{\It}{\mathcal{I}(t)}

\begin{document}
\vspace*{0.2in}

\begin{flushleft}
{\Large
\textbf\newline{A Ca\textsuperscript{2+} puff model based on integrodifferential equations} 
}
\newline
\\
Molly Hawker\textsuperscript{1},
Pengxing Cao\textsuperscript{5},
James Sneyd\textsuperscript{6},
Ivo Siekmann\textsuperscript{1 2 3 4}
\\
\bigskip
\textbf{1}  School of Computer Science and Mathematics, Faculty of Engineering and Technology, James Parsons Building, Liverpool John Moores University, 3 Byrom Street, Liverpool, Merseyside L3 3AF, UK
\\
\textbf{2} Liverpool Centre for Cardiovascular Science, Faculty of Health and Life Sciences
University of Liverpool, Foundation Building, Brownlow Hill, Liverpool, L69 7TX
\\
\textbf{3} Data Science Research Centre, Liverpool John Moores University
\\
\textbf{4} Protect e-Health group, Liverpool John Moores University
\\
\textbf{5} School of Mathematics and Statistics, University of Melbourne, Melbourne, VIC, Australia
\\
\textbf{6} Department of Mathematics, University of Auckland, Auckland, New Zealand
\bigskip

%
%





* m.a.hawker@2021.ljmu.ac.uk

\end{flushleft}
\section*{Abstract}
The calcium signalling system is important for many cellular processes within the human body. 
Signals are transmitted within the cell by releasing calcium (\ca) from the endoplasmic reticulum (ER) into the cytosol via clusters of \ca\ channels. Mathematical models of \ca\ release via inositol 1,4,5-trisphosphate receptors (\ipr) help with understanding underlying \ca\ dynamics but data-driven modelling of stochastic calcium release events, known as \ca\ puffs, is a difficult challenge. Parameterising Markov models for representing the \ipr\ with steady-state single channel data obtained at fixed combinations of the ligands \ca\ and inositol-trisphosphate (\ipthree) has previously been demonstrated to be insufficient
. However, by extending an \ipr\ model based on steady-state data with an integral term that incorporates the delayed response of the channel to varying \ca\ concentrations we succeed in generating realistic \ca\ puffs. By interpreting the integral term as a weighted average of \ca\ concentrations that extend over a time interval of length~$\tau$ into the past we conclude that the \ipr\ requires a certain amount of memory of past ligand concentrations. 


\section*{Author summary} 
The release of calcium ions by ion channels is important for many human biological processes. Abnormal ion channels can cause dysfunction in the release of calcium ions, and thus be detrimental to an individual’s health. Hybrid stochastic systems are often used to model the dynamics of the ion channel and calcium release, with Markov models simulating the stochastic behaviour of the ion channel and ordinary differential equations modelling the deterministic release of calcium ions. We propose a new architecture for modelling the inositol-trisphosphate receptor (\ipr) that makes it easy to integrate steady state data 
recorded at constant ligand concentrations with data sets that give insight into the delayed response of the channel to changes in ligand concentrations. The dependency of our \ipr\ model on calcium (\ca) can be interpreted as a weighted average of \ca\ concentrations in the past which indicates that the channel requires a certain amount of ``memory'' to function properly. Thus, apart from the fact that our model can be easily parametrised in a modular fashion with different types of single channel data, it also provides interesting insights into the biophysics of the \ipr.


\section*{Introduction}
The calcium (Ca\textsuperscript{2+}) signalling system is vital for cellular function, playing an important role in both excitable and non-excitable cells. This includes contracting and relaxing cardiomyocytes, controlling many psychological processes and regulating several major ion flux mechanisms \cite{Fearnley2011, Calì2014a, Wagner2012,Garcia2017,Han2017, Glaser2019}. However, the Ca\textsuperscript{2+} signalling system is not infallible and has been linked to numerous human disease states, such as hypertrophy, congestive heart failure, neurological diseases and the inhibition of salivary secretion \cite{ Berridge1997,Tveito2016, Han2017, Glaser2019}. Therefore, it is important to understand Ca\textsuperscript{2+} dynamics further, and this can be achieved through mathematical modelling.

Inositol 1,4,5-trisphosphate receptors (IP\textsubscript{3}Rs) are located in the membrane of the endoplasmic reticulum and sarcoplasmic reticulum and regulate the release of \ca\ ions by opening and closing stochastically \cite{Berridge1997,Bootman2012}. \ipr\ are distributed across the cell in clusters of unknown size \cite{Swillens1994}. The concentration of Ca\textsuperscript{2+} released from a cluster of IP\textsubscript{3}Rs can be described in a hierarchical manner \cite{Yao1995,Berridge1997,Bootman1997,Marchant2001,Rudiger2007, Smith2009, Ruckl2015}. The binding of inositol 1,4,5-trisphosphate (IP\textsubscript{3}) to an activating site of an IP\textsubscript{3}R opens the IP\textsubscript{3}R, releasing Ca\textsuperscript{2+} ions into the cytoplasm \cite{Berridge2000,Bootman2012}. This small increase in the cytoplasmic Ca\textsuperscript{2+} concentration is known as a Ca\textsuperscript{2+} blip. The release of Ca\textsuperscript{2+} ions from a Ca\textsuperscript{2+} blip stimulates neighbouring IP\textsubscript{3}-liganded IP\textsubscript{3}Rs, increasing their open probability and releasing further Ca\textsuperscript{2+} ions into the cytoplasm \cite{Foskett2007, Skupin2010 ,Rudiger2019, Siekmann2019}. Ca\textsuperscript{2+} released from a cluster of IP\textsubscript{3}Rs is called a Ca\textsuperscript{2+} puff. The occurrence of many Ca\textsuperscript{2+} puffs can trigger a wave of Ca\textsuperscript{2+} across the entire cell \cite{Bezprozvanny1991, Berridge1997, Marchant2001}. A high concentration of Ca\textsuperscript{2+} decreases the open probability of the IP\textsubscript{3}R and inhibits the channel \cite{Bezprozvanny1991,Mak2007,Siekmann2019}. Intracellular oscillations and waves are important cellular signals; Ca\textsuperscript{2+} puffs are believed to play a vital role in generating the Ca\textsuperscript{2+} waves that travel across the cell \cite{Bootman1997, Marchant2001, Ruckl2015}.

In order to understand Ca\textsuperscript{2+} dynamics, mathematical models of the IP\textsubscript{3}R have been developed \cite{Keizer1994, Li1994, Swillens1994,Sneyd2004,Siekmann2012, Ullah2012, Cao2013, Rudiger2013, Cao2014, Dupont2016, Dupont2017, Han2017, Siekmann2019}. An early example of continuous-time Markov chains being used to analyse ion channel behaviour can be found in \cite{Colquhoun1977}. Over the past decade it has become evident that parameterising continuous-time Markov models using experimental data, whilst challenging, leads to more accurate simulations \cite{Siekmann2019}. The model by Siekmann et al. \cite{Siekmann2012} incorporates the large single-channel data set by Wagner and Yule \cite{Wagner2012} and accurately accounts for modal gating of the IP\textsubscript{3}R i.e. the spontaneous switching between a high and a low level of activity. 
Cao et al. \cite{Cao2013} observed that in its original form, the Siekmann et al. model \cite{Siekmann2012} could not be used for accurately simulating \ca\ puffs. They hypothesised that this was due to the fact that the model has been parametrised by steady-state data \cite{Wagner2012} that had been obtained from experiments where \ca\ concentrations were held constant. By integrating data by Mak et al. \cite{Mak2007} they developed an extension of the model which they used successfully for simulating realistic puff distributions. In contrast to the steady-state data by Wagner and Yule \cite{Wagner2012}, the Mak et al. data set \cite{Mak2007} behaviour of the \ipr\ when \ca\ concentrations vary in time---they observed a
delayed response of the IP\textsubscript{3}R to rapid changes in Ca\textsuperscript{2+} concentrations.
Further developments of the model have since been made, such as creating a two-state model by using a quasi-steady-state approximation that removes states with short dwell times that account for very brief openings and closings, simulating the dynamics in HSY cells and understanding the dependencies of certain parameters on the interpuff interval (the waiting time between subsequent puffs) \cite{Cao2014,Han2017,Cao2017}.



The goal of the study presented here is to develop a model for the \ipr\ that accurately accounts for the delayed response to changes in \ca\ concentration as described by Mak et al. \cite{Mak2007}. To detect changes in the \ca\ concentration $c(t)$ over a period of time, rather than just ``sensing''~$c(t)$ at the current time~$t$ the \ipr\ must ``observe'' the \ca\ concentrations over a time interval~$\It =[t-\tau, t]$ that reaches a certain length of time~$\tau$ in the past. 

We introduce an integral over the \ca\ concentration~$c(t)$ over the time interval~$\It$:
\begin{equation}
    \label{eq:avgCa}
    \bar{c}(t)=\frac{1}{\tau} \int_{t-\tau}^t f(c(s)) ds    
\end{equation}

with~$f:\mathbb{R}^+ \to \mathbb{R}^+$ and $\tau>0$. For~$\tau=0$ we set~$\bar{c}(t)=c(t)$. Choosing $f=\text{id}$ i.e. $\bar{c}(t)=\frac{1}{\tau} \int_{\tau-t}^t c(s) ds$ Eq \eqref{eq:avgCa} is a temporal average of~$c(t)$ over the interval~$\It$, thus, for general positive $f$, Eq \eqref{eq:avgCa} can be interpreted as a weighted temporal average of~$c(t)$ over~$\It$.

Models such as the Siekmann et al. model \cite{Siekmann2012} are formulated as \ca-dependent infinitesimal generators~$Q(c)$; by allowing for time-dependent~$c(t)$ we obtain the time-dependent infinitesimal generator~$Q(c(t))$. In order to account for the delayed response to changes in \ca\ concentrations \cite{Mak2007} we propose instead of substituting the current \ca\ concentration~$c(t)$ at time~$t$ into~$Q(c)$, we replace~$c$ with the weighted temporal average~$\bar{c}(t)$ defined by \eqref{eq:avgCa} i.e. the time-dependent infinitesimal generator of our new model is~$Q(\bar{c}(t))$ rather than~$Q(c(t))$. Moreover, because for~$\tau=0$, we have~$Q(\bar{c}(t))=Q(c(t))$, using the temporal average~$\bar{c}(t)$ defined by \eqref{eq:avgCa} is a natural extension of defining the time-dependent infinitesimal generator using the current \ca\ concentration~$c(t)$. 

This shows that for a given infinitesimal generator which has been parametrised using data obtained at constant \ca\ concentrations such as Wagner and Yule \cite{Wagner2012}, a model that appropriately responds to changes in \ca\ concentration can be found by determining the time~$\tau$ and the function~$f$ in order to obtain $\bar{c}(t)$ \eqref{eq:avgCa}. In the current study, rather than searching for suitable functions~$f$ we will take advantage of the fact that our new model can, in fact, be shown to be mathematically equivalent to the model by Cao et al. \cite{Cao2013} if we choose~$\tau=\infty$. In their extension of the Siekmann et al. model, Cao et al. \cite{Cao2013} introduced---in addition to the system of equations representing the Markov model based on the generator~$Q$---four gating variables, analogous to those in the classical Hodgkin-Huxley model \cite{Hodgkin1952}. Following an approach demonstrated by Brady \cite{Brady1972} for the Hodgkin-Huxley model, it is possible to represent the gating variables as integrals by explicitly solving the four linear differential equations for the gating variables. Thus, by incorporating these integrals into the infinitesimal generator~$Q(c)$ we obtain a system of integrodifferential equations with infinitesimal generator~$Q(\bar{c}(t))$ which is equivalent to the model by Cao et al. \cite{Cao2013} if we choose~$\tau=\infty$.

This brings us to the interpretation of the parameter~$\tau$. Considering that~$\tau$ determines the length of the interval~$\It$ used for averaging the \ca\ concentration before the current time~$t$, it is natural to relate~$\tau$ to how long the \emph{memory} of \ca\ concentrations reaches into the past. Following this interpretation the model by Cao et al. \cite{Cao2013} corresponds to the idealised situation that the \ipr\ has infinite memory. Developing our model based on the model by Cao et al. \cite{Cao2013} enables us to investigate the effect of reducing the time~$\tau$ that the memory of the channel reaches into the past to a more realistic value. Moreover, more importantly, varying~$\tau$ allows us to explore how much memory is needed for the model to produce realistic puffs.


At first glance, it might seem counter-intuitive or even inappropriate to introduce the notion of memory in the framework of Markov models, considering that the Markov property is often associated with ``memorylessness'' i.e. the absence of memory. However, the ``memorylessness'' of Markov models is, in fact, not as strict as it may seem. The class of Markov models which are commonly used for representing ion channels---examples include \cite{Colquhoun1981,Siekmann2012,Ullah2012,Cao2013,Tveito2016,Dupont2016,Friedhoff2021}---are so-called aggregated Markov models \cite{Fre:86a} which consist of multiple open and closed states. In fact, the Markov property implies ``memorylessness'' only for individual states of a Markov model. The sojourn time in a single Markov state~$S$ follows an exponential distribution. The exponential distribution is the only continuous distribution which is ``memoryless''. This means that the probability for a sojourn time larger than $T$ is not affected by how much time~$T_0$ has already been spent in~$S$, or formally,  $\prob(t > T + T_0 | T_0)=\prob(t>T)$. This does not hold for aggregates of states for which the sojourn time distribution is a mixture of exponential distributions. For example, if a model has one ``fast'' state~$S_{\text{fast}}$ and one ``slow'' state $S_{\text{slow}}$, the longer it takes for a sojourn in the aggregate $\mathcal{S}=\{S_{\text{fast}}, S_{\text{slow}} \}$ to end, the more likely it is that we will observe a long sojourn time that is characteristic of the slow state $S_{\text{slow}}$. This can be interpreted as the channel having ``memory'' because it ``remembers'' how much time has already been spent in the aggregate~$\mathcal{S}$ \cite{Galea2017, Galea2020,Silva2021}. 

Rather than ``memory'' of the time spent in the aggregate of open or closed states, via the dependency on the weighted average $\bar{c}(t)$ over \ca\ concentrations ranging from the current time~$t$ until a length of time~$\tau$ in the past, we introduce ``memory'' of past \ca\ concentrations. Thus, the main hypothesis of our study is that the delayed response of the \ipr\ to changes in \ca\ concentrations seen in the experiments by Mak et al. \cite{Mak2007} can be explained by how much \ca\ the channel has been exposed to; not just at the current time~$t$ but in the recent past. We defer possible interpretations of this memory effect to the Discussion.


In this article, we produce a Ca\textsuperscript{2+} puff model based on producing a Ca\textsuperscript{2+} puff model based on integrodifferential equations. Firstly, we demonstrate that the IP\textsubscript{3}R requires ``knowledge'' of past Ca\textsuperscript{2+} concentrations in order to produce Ca\textsuperscript{2+} puffs. Next, we show the gating variables used in the Markov model by Cao et al. \cite{Cao2013} can be directly replaced with integrals. We use quasi-steady state approximation to reduce the model to consist of six states and one gating variable. Finally, our model is simplified further to a two-state model with one gating variable. We show that a six-state model with four gating variables can be transformed to a system of two integrodifferential equations. 

\section*{Materials and methods}
\subsection*{An integrodifferential version of the Siekmann model}
\begin{figure}
    \centering
    \includegraphics[scale=0.7]{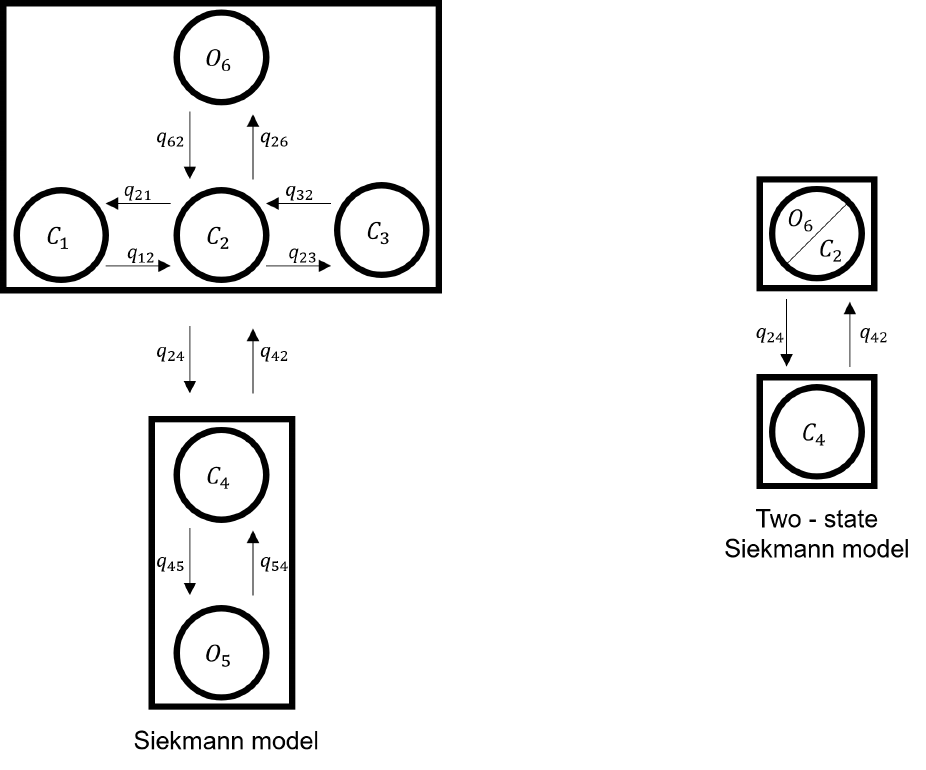}
    \caption{The structure of the six-state and two-state Siekmann Model. The six-state model \cite{Siekmann2012,Cao2013}: the active mode consists of states C\textsubscript{1}, C\textsubscript{2}, C\textsubscript{3} and O\textsubscript{6}; the inactive mode consists of states C\textsubscript{4} and O\textsubscript{5}. The two-state model \cite{Cao2014}: The active mode consists of the joint states C\textsubscript{2} and O\textsubscript{6}; the inactive mode consists of the closed state C\textsubscript{4}.}
    \label{Fig1}
\end{figure}

The Siekmann model is a six-state Markov model, with four closed states and two open states \cite{Siekmann2012}. The model, shown on the left in Figure \ref{Fig1}, has two modes. The first mode consists of four states and the second mode of two states. These modes describe the open probability of the ion channel. When the channel is in the four-state mode, known as the active mode, it has an open probability of around \num{0.7}, whereas when the channel is in the two-state mode, known as the inactive mode, it has an open probability of around \num{0}. All the transition rates between the states are constant with the exception of q\textsubscript{24} and q\textsubscript{42} which are both Ca\textsuperscript{2+} and IP\textsubscript{3} dependent.

The differential equations describing the transitions between states can be represented in matrix form, with a matrix of the transition rates and a vector of the states, known as the Q matrix. The Q matrix for the six-state Siekmann model is presented in Eq \ref{eq:6-state matrix}.

\begin{equation}
\begin{Bmatrix}
    \frac{dC_{1}}{dt}\\
\frac{dC_{2}}{dt}\\
    \frac{dC_{3}}{dt}\\
    \frac{dC_{4}}{dt}\\
    \frac{dO_{5}}{dt}\\
    \frac{dO_{6}}{dt}
\end{Bmatrix}
=
\begin{Bmatrix}
\centering
-\text{q}_{12} & \text{q}_{12} & 0 & 0 & 0 & 0\\
\text{q}_{21} &  -(\text{q}_{21}+\text{q}_{23}+\text{q}_{24}+\text{q}_{26}) & \text{q}_{23} & \text{q}_{24} & 0 & \text{q}_{26}\\
0 & \text{q}_{32} & -\text{q}_{32} & 0 & 0 & 0 \\
0 & \text{q}_{42} & 0 & -(\text{q}_{42}+\text{q}_{45}) & \text{q}_{45} & 0 \\
0 & 0 & 0 & \text{q}_{54} & -\text{q}_{54} & 0 \\
0 & \text{q}_{62} & 0 & 0 & 0 & -\text{q}_{62}
\end{Bmatrix}
\begin{Bmatrix}
    C_{1}\\
    C_{2}\\
    C_{3}\\
    C_{4}\\
    O_{5}\\
    O_{6}
\end{Bmatrix}
\label{eq:6-state matrix}
\end{equation}

The rates q\textsubscript{24} and q\textsubscript{42} are calculated using two 
\ca-dependent variables each, $m_{24}$, $h_{24}$, $m_{42}$, $h_{42}$ as shown in Eq \eqref{eq:q24} and \eqref{eq:q42}. The parameters a\textsubscript{24}, a\textsubscript{42}, V\textsubscript{24} and V\textsubscript{42} are constant.

\begin{align}
    q_{24}&=a_{24}+V_{24}(1-m_{24}h_{24}) \label{eq:q24} \\
    q_{42}&=a_{42}+V_{42} m_{42} h_{42} \label{eq:q42}
\end{align}

If we replace $m_{24}$, $h_{24}$, $m_{42}$, $h_{42}$ with the \ca-dependent $m_{24 \infty}$, $h_{24 \infty}$, $m_{42 \infty}$, $h_{42 \infty}$ defined as follows:
\begin{align}
    m_{24 \infty} &= \frac{c^{n_{24}}}{c^{n_{24}}+k_{24}^{n_{24}}}, \label{eq: m24} \\
    h_{24 \infty} &= \frac{k_{-24}^{n_{-24}}}{c^{n_{-24}}+k_{-24}^{n_{-24}}}, \label{eq: h24} \\
    m_{42 \infty} & = \frac{c^{n_{42}}}{c^{n_{42}}+k_{42}^{n_{42}}}, \label{eq: m42} \\
    h_{42 \infty} & = \frac{k_{-42}^{n_{-42}}}{c^{n_{-42}}+k_{-42}^{n_{-42}}}, \label{eq: h42}
\end{align}
the resulting rates~$q_{24}$ and~$q_{42}$ fit the \ca\ dependency of these rates inferred by Siekmann et al. \cite{Siekmann2012} from the data by Wagner \& Yule. \cite{Wagner2012}.

\subsubsection*{The Cao et al. model}

 Cao et al. \cite{Cao2013} observed that the model~$Q(c)$~\eqref{eq:6-state matrix} with the \ca-dependent rates~$q_{24}$ and~$q_{42}$, \eqref{eq:q24} and \eqref{eq:q42}, parametrised by~\eqref{eq: m24}-\eqref{eq: h42} failed to produce realistic puffs. 
They decided to introduce a delayed response to changes in the \ca\ concentration~$c(t)$ by representing~$m_{24}$, $m_{42}$, $h_{24}$ and $h_{42}$ as Hodgkin-Huxley-like gating variables\cite{Hodgkin1952}:

\begin{equation}
    \frac{dG}{dt}=\lambda_{G}(G_{\infty}-G) \label{eq:gv ODES}
\end{equation}

where $G=m_{24},m_{42},h_{24},h_{42}$ and $G_{\infty}=m_{24 \infty}, h_{24 \infty}, m_{42 \infty}, h_{42 \infty}$. In the original Siekmann model, in response to a change in the \ca\ concentration, the variables~$G$ are immediately set to~$G_{\infty}$. In contrast, when modelling~$G$ as gating variables~\eqref{eq:gv ODES}, rather than instantaneously attaining~$G_{\infty}$, a variable~$G$ instead approaches~$G_{\infty}$ from its current value at the rate~$\lambda_{G}$.








The rates at which $m_{24}$, $h_{24}$ and $m_{42}$ reach their equilibrium are constant \cite{Cao2013}. However, $h_{42}$ has a more complex dynamic and its rate was modelled heuristically by Cao et al. \cite{Cao2013} as:

\begin{equation}
    \lambda_{h_{42}}=a_{h_{42}}+\frac{V_{h_{42}}c^7}{c^7+20^7} \label{eq: h42 rate}
\end{equation}

Where $\text{a}_{\text{h}_{42}}$ and V\textsubscript{$h_{42}$} are constants. When the Ca\textsuperscript{2+} concentration is low, the rate $\lambda$\textsubscript{$h_{42}$} will be low. Similarly, when the Ca\textsuperscript{2+} concentration is high, $\lambda$\textsubscript{$h_{42}$} will be high. The parameters of the gating variable equations were chosen so that the resulting model showed a delayed response consistent with the Mak et al. \cite{Mak2007} data.


\subsubsection*{ Calculating $\bar{c}(t)$ from the Cao et al. model}

As explained in the Introduction, our aim is to find a suitable weighted average~$\bar{c}(t)$ so that the resulting model~$Q(\bar{c}(t))$ exhibits a delay in response to change in the \ca\ concentration as observed by Mak et al. \cite{Mak2007} which is essential in producing realistic puffs. We will see that it is possible to find an expression for the averaged \ca\ concentration~$\bar{c}(t)$ proposed in~\eqref{eq:avgCa} by integrating the differential equations for the gating variables~\eqref{eq:gv ODES} as demonstrated by Brady \cite{Brady1972} for the Hodgkin-Huxley equations \cite{Hodgkin1952}.


For each gating variable~$G$ we obtain the integral expression 
\begin{align}
    \Phi_{G}(t,c)&=G(0)\exp\left[-\int_{0}^{t}(\lambda_{G} \circ c)(x) dx\right] \nonumber\\
    &- \exp\left[- \int_{0}^{t}(\lambda_{G} \circ c)(x) dx\right] 
    \int_{0}^{t}(-\alpha_{G} \circ c)(s) 
    \cdot \exp\left[\int_{0}^{s}(\lambda_{G} \circ c)(x) dx\right]ds \nonumber \\
    &=\exp(-J(t)) \left[ G(0) + \int_{0}^{t}(\alpha_{G} \circ c)(s) \cdot \exp (J(s))ds \right]
\label{eq:Brady1972 lambda}
\end{align}

where $G$ represents the gating variable and $c$ the Ca\textsuperscript{2+} concentration and

\begin{equation}
    \label{eq:intLambdaC}
    J(t) = \int_{0}^{t}(\lambda_{G} \circ c)(x) dx.
\end{equation}

The initial values are: $t_0=0$, $c(t_{0})=\num{0.1}$ \unit{\micro \Molar}, $G(t_{0})=\frac{\alpha_{G}(0)}{\lambda_{G}(0)}$.

$\alpha$\textsubscript{G} is calculated as follows, using rates presented in Table \ref{table: parameters}:

\begin{equation}
    \alpha_{G} = \lambda_{G}G_{\infty}
    \label{eq: alpha}
\end{equation} 

We will now verify if the~$\Phi_G(t,c)$ calculated in Eq~\eqref{eq:Brady1972 lambda} are indeed functions of an appropriately defined weighted average~$\bar{c}(t)$ as introduced in Eq.~\eqref{eq:avgCa}. It is easy to see that $J(t)$ is of the required form by simply factoring out $1/\tau$. The term $\int_{0}^{t}(\alpha_{G} \circ c)(s) \cdot \exp (J(s))ds$ can be interpreted as a different version of a weighted average $\bar{c}(t)$---the function~$\exp (J(s))ds\cdot(\alpha_{G} \circ c)(s)$ is a positive function applied to $c$. This shows that $\Phi_G$ is a function of two different weighted averages~$\bar{c}(t)$. 

In principle, the transformation to integrodifferential equations introduces an infinite delay i.e. the integrals replacing the gating variables extend over the time interval $(-\infty, t]$. This not only makes the numerical solution of the model equations computationally infeasible but also implies that the IP$_3$R has an ``infinite'' memory which appears unrealistic. For this reason, we consider integrals with finite delays,~$\tau$ (see Eq \eqref{eq:Brady1972 G finite}). $\tau$ can be interpreted as how far into the past the ion channel's memory spans.

The general version of the finite time integral term can be written as:

\begin{equation}
    \Phi_{G}(t,c)=\exp(-J(t)) \left[ G(0) + \int_{0}^{t}(\alpha_{G} \circ c)(s) \cdot \exp (J(s))ds \right]
\label{eq:Brady1972 G finite}
\end{equation}

We replace the ODEs in the Siekmann model (Eq \eqref{eq:gv ODES}) with the integrodifferential equation described in Eq \eqref{eq:Brady1972 G finite}.

\begin{table}
\small
   \caption{Model parameters. IP\textsubscript{3}-dependent parameters are evaluated at a concentration of \num{0.1} \unit{\micro \Molar} as indicated by subscripts. Full model details are given in \cite{Cao2013}.}
\vspace{0.1cm}
\begin{tabularx} {\textwidth}[t]{l X r l}
    \hline
    Symbol      & Description           & Value       & Units\\
    \hline
    \multicolumn{4}{c}{\textsl{Gating kinetics}}\\
\hline
  $a_{24}$ & Basal level of $q_{24}$ & 29.85$_{p=\unit{0.1}\unit{\micro \Molar}}$ & $\unit{\second}^{-1}$ 
\\
$V_{24}$ & Gating-dependent part of~$q_{24}$ & 312.85$_{p=\unit{0.1}\unit{\micro \Molar}}$ & $\unit{\second}^{-1}$ 
\\
$a_{42}$ & Basal level of $q_{42}$ & 0.05$_{p=\unit{0.1}\unit{\micro \Molar}}$ & $\unit{\second}^{-1}$ 
\\
$V_{42}$ & Gating-dependent part of~$q_{42}$ & 100 & $\unit{\second}^{-1}$
\\
$\lambda_{h_{24}}$ & Rate of approach to steady state of~$h_{24}$ & 40 & $\unit{\second}^{-1}$
\\
$n_{-24}$ & Hill coefficient for $\text{Ca}^{2+}$\ dependency of~${h_{24}}_{\infty}$ & 0.04$_{p=\unit{0.1}\unit{\micro \Molar}}$  

\\
  $k_{-24}$ & Half-saturation constant for  $\text{Ca}^{2+}$\ dependency of~${h_{24}}_{\infty}$ & 97.00$_{p=\unit{0.1}\unit{\micro \Molar}}$ & 
 \medskip \\ \medskip
  ${h_{24}}_{\infty}$  & Steady state of~$h_{24}$ & $\dfrac{k_{-24}^{n_{-24}}}{c^{n_{-24}}+k_{-24}^{n_{-24}}}$ \medskip \\ \medskip
$a_{h_{42}}$ & Basal level of~ $\lambda_{h_{42}}$ (tuning parameter) & 0.5 & $\unit{\second}^{-1}$ \\
$V_{h_{42}}$ &  $\text{Ca}^{2+}$-dependent part of~$\lambda_{h_{42}}$ & 100 & $\unit{\second}^{-1}$ \\
$K_{h_{42}}$ & Half-saturation constant for  $\text{Ca}^{2+}$-dependency of~$\lambda_{h_{42}}$ & 20 & \unit{\micro \Molar} \\
  $\lambda_{h_{42}}$ & Rate of approach to steady state of~$h_{42}$ & $a_{h_{42}}+\dfrac{V_{h_{42}}c^7}{c^7+K_{h_{42}}^7}$ & $\unit{\second}^{-1}$  \medskip \\ \medskip
  $n_{-42}$ & Hill coefficient for  $\text{Ca}^{2+}$\ dependency of~${h_{42}}_{\infty}$ & 3.23$_{p=\unit{0.1}\unit{\micro \Molar}}$ 
\\
  $k_{-42}$ & Half-saturation constant for  $\text{Ca}^{2+}$\ dependency of~${h_{42}}_{\infty}$ & 0.17$_{p=\unit{0.1}\unit{\micro \Molar}}$
& 
 \medskip \\\medskip
  ${h_{42}}_{\infty}$ & Steady state of~$h_{42}$  & $\dfrac{k_{-42}^{n_{-42}}}{c^{n_{-42}}+k_{-42}^{n_{-42}}}$ \medskip \\\medskip
  $\lambda_{m_{24}}$ & Rate of approach to steady state of~$m_{24}$ & 100 &  $\unit{\second}^{-1}$  \\
  $n_{24}$ & Hill coefficient for  $\text{Ca}^{2+}$\ dependency of~${m_{24}}_{\infty}$ & 6.31$_{p=\unit{0.1}\unit{\micro \Molar}}$
\\
  $k_{24}$ & Half-saturation constant for  $\text{Ca}^{2+}$\ dependency of~${m_{24}}_{\infty}$ & 0.549$_{p=\unit{0.1}\unit{\micro \Molar}}$ 
\medskip \\ \medskip 
  ${m_{24}}_{\infty}$ & Steady state of~$m_{24}$ & $\dfrac{c^{n_{24}}}{c^{n_{24}}+k_{24}^{n_{24}}}$&  \medskip
  \\
  \medskip 
  $\lambda_{m_{42}}$ & Rate of approach to steady state of~$m_{42}$ & 100  & $\unit{\second}^{-1}$\\
  $n_{42}$ & Hill coefficient for  $\text{Ca}^{2+}$\ dependency of~${m_{42}}_{\infty}$ & 11.16$_{p=\unit{0.1}\unit{\micro \Molar}}$
\\
$k_{42}$ & Half-saturation constant for  $\text{Ca}^{2+}$\ dependency of~${m_{42}}_{\infty}$ & 0.40$_{p=\unit{0.1}\unit{\micro \Molar}}$ & 
\medskip \\ \medskip 
  ${m_{42}}_{\infty}$ & Steady state of~$m_{42}$ & $\dfrac{c^{n_{42}}}{c^{n_{42}}+k_{42}^{n_{42}}}$ \\
\hline
\multicolumn{4}{c}{\textsl{$\text{Ca}^{2+}$\ balance}}\\
\hline
  $c_h$ & Elevated $\text{Ca}^{2+}$\ in vicinity of open \ipr\ channel & 120 & $\unit{\micro \Molar}$  \\
  $B$ & Total buffer concentration &20 & $\unit{\micro \Molar}$ \\
  $\kOn$  & Binding of fluo4 buffer to \ca  & 150 & $\unit{\micro \Molar \unit{\second}^{-1}}$ \\
  $\kOff$  & Unbinding of fluo4 buffer from \ca & 300 & $\unit{\second}^{-1}$ \\
  $\Jr$ & Flux of $\text{Ca}^{2+}$\ through single channel & 200 & $\unit{\micro \Molar \unit{\second}^{-1}}$\\
  $\Jleak$ & $\text{Ca}^{2+}$\ influx from cluster environment & 33 & $\unit{\micro \Molar \unit{\second}^{-1}}$\\
  $V_d$ & Rate of cytoplasmic $\text{Ca}^{2+}$\ removal from the cluster   &  4000& $\unit{\micro \Molar}$ $\unit{\second}^{-1}$\\
  $K_d$  & Half-saturation constant for cytoplasmic $\text{Ca}^{2+}$\ removal  & 12 & $\unit{\micro \Molar}$ \\
    \hline
\end{tabularx}
 \label{table: parameters}
\end{table}

\subsection*{The reduced IP\textsubscript{3}R model} 
Quasi-steady-state approximation replaces the ODEs for fast variables with their steady state. This reduces the number of equations in the system, leaving only a system for slow variables \cite{Vejchodský2014}. Cao et al. and Dupont et al. \cite{Cao2014, Dupont2016} state the rate at which the gating variables $m_{24}$, $h_{24}$ and $m_{42}$ reach their steady state is so quick, they can be set equal to their steady state.

That is:

\begin{align}
    m_{24} = m_{24 \infty}, \hspace{0.5cm} h_{24} = h_{24 \infty}, \hspace{0.5cm} m_{42} = m_{42 \infty}
\end{align}

Whilst, the reduced model still consists of six integrodifferential equations, computationally it is simpler because it only uses one gating variable.

\subsection*{The reduced two-state IP\textsubscript{3}R model} 
Cao et al. \cite{Cao2014} showed the six-state IP\textsubscript{3}R model can be reduced to a two-state IP\textsubscript{3}R model without qualitatively changing the Ca\textsuperscript{2+} puff dynamics.

The right schematic presented in Figure \ref{Fig1} describes the two-state model by Cao et al. \cite{Cao2014}. In this model, only the inter-modal transitions have an effect on IP\textsubscript{3}R behaviour and the structure of the active and inactive modes seen within the six-state model are ignored \cite{Cao2014}. Constant parameters for rates q\textsubscript{24} and q\textsubscript{42} remain the same as those in Eq \eqref{eq:q24} and \eqref{eq:q42}. Due to the reduction in the model, q\textsubscript{24} is scaled by $\frac{\text{q}_{26}}{\text{q}_{62}+\text{q}_{26}}$. See \cite{Cao2013} for details.

\subsection*{Deterministic calcium dynamics}
Using the same system of ODEs as in \cite{Cao2013}, we develop a model that accounts for various fluxes that influence the Ca\textsuperscript{2+} concentration, $c$, in the cytosol as well as the Ca\textsuperscript{2+} dye, $b_{\text{fluo4}}$.

\begin{align}
    \frac{dc}{dt} &= J_{\text{increase}} N_{\text{o}}+ J_{\text{leak}} - J_{\text{decrease}}-k_{\text{on}}(B_{\text{fluo4}}-b_{\text{fluo4}})c+k_{\text{off}} b_{\text{fluo4}} \label{eq:Calcium flux} \\
    \frac{db_{\text{fluo4}}}{dt}&=k_{\text{on}}(B_{\text{fluo4}}-b_{\text{fluo4}} )c-k_{\text{off}} b_{\text{fluo4}} \label{eq:Calcium dye}
\end{align}

Ca\textsuperscript{2+} fluxes can be modelled deterministically using ODEs. Eq \eqref{eq:Calcium flux} and \eqref{eq:Calcium dye} describe the Ca\textsuperscript{2+} concentration and the Ca\textsuperscript{2+} dye, respectively. The Ca\textsuperscript{2+} flux is described through the parameters $J_{\text{increase}}$, $J_{\text{leak}}$ and $J_{\text{decrease}}$. $J_\text{increase}$ represents the Ca\textsuperscript{2+} flux through an open IP\textsubscript{3}R; $N_{\text{o}}$ is the number of open IP\textsubscript{3}R in a cluster. The leakage of Ca\textsuperscript{2+} from the endoplasmic reticulum is described using $J_{\text{leak}}$. $J_{\text{decrease}}$ represents the Ca\textsuperscript{2+} flux that returns to the endoplasmic reticulum \cite{Cao2013, Siekmann2019}. 
Eq \eqref{eq:Calcium dye} represents the Ca\textsuperscript{2+} dye bound to Ca\textsuperscript{2+} that is detected using a light microscope within experiments \cite{Cao2013}. The changes in the Ca\textsuperscript{2+} signalling can be visualised through the changes in the fluorescence light correlating with changes in Ca\textsuperscript{2+} signalling \cite{Pratt2020}. This process is described in Eq \eqref{eq:Calcium flux}, \eqref{eq:Calcium dye} using parameters $B_{\text{fluo4}}$ and $b_{\text{fluo4}}$, which represent the total dye buffer concentration and the Ca\textsuperscript{2+}-bound dye buffer concentration, respectively \cite{Siekmann2019}. For the two-state model, all parameters remain the same as those for the six-state model with the exception of J\textsubscript{increase} which is replaced with J\textsubscript{increase} $\cdot$ $\frac{q_{26}}{(q_{62}+q_{26})}$ \cite{Cao2014}. Parameter values are detailed in Table \ref{table: parameters}.

\subsection*{Calcium puff statistics}
Ca\textsuperscript{2+} puffs are often characterised by taking into consideration three key statistics: the interpuff interval (IPI), the puff amplitude and the puff duration. IPIs are defined as being the time between the peak amplitude of Ca\textsuperscript{2+} puffs. We determine the start of a Ca\textsuperscript{2+} puff as being when the Ca\textsuperscript{2+} concentration is 20\% of the peak amplitude. Similarly, the end of the puff is calculated the time after the peak where the Ca\textsuperscript{2+} concentration is 20\% of the peak amplitude. The difference in the end and start times determines the duration of the Ca\textsuperscript{2+} puff.

 Thurley et al. \cite{Thurley2011a} proposed a time-dependent variant of the exponential distribution for fitting to IPI data. We fit our simulated IPI distributions to this probability density function by calculating the suitable parameters for it. The time-dependent distribution is:

\begin{equation}
    P_{IPI}=\lambda(1-\exp{(-\xi t)})\exp{(-\lambda t +\lambda(1-\exp{(-\xi t}))/\xi)}
 \label{eq:time dependent distribution}
\end{equation}

where $\lambda$ is the puff rate and $\xi$ is the recovery rate. We estimated the mean IPI from the data and set $\lambda$ as the reciprocal of this value, as previously demonstrated by Cao et al. \cite{Cao2017}. $\xi$ is optimised using the \textit{lsqcurvefit} function in Matlab. 

\subsection*{Numerical methods}
We solve Eq \eqref{eq:Calcium flux}, \eqref{eq:Calcium dye} using the fourth-order Runge-Kutta method. The dynamics of the Markov models representing the IP$_3$R channels is simulated with a Gillespie algorithm. Due to the rates q\textsubscript{24} and q\textsubscript{42} being Ca\textsuperscript{2+} dependent, they are time-dependent. For this reason the original Gillespie algorithm cannot be used. Adaptive timing, as detailed in \cite{Alfonsi2005,Cao2013,Rudiger2013}, is used to make the algorithm more run-time-efficient. A maximum time step size of \num{10}\textsuperscript{-4} \unit{\second} is used for the six and two-state models. Integrals in Eq \eqref{eq:Brady1972 G finite} are calculated using the Riemann Sum, using a larger time step (\num{10}\textsuperscript{-2} \unit{\second}). As evidenced in \nameref{S3_Fig}, the increased time-step strongly increases computational efficiency whilst not significantly decreasing the approximation accuracy of the integral. IP\textsubscript{3} is set to \num{0.1} \unit{\micro \Molar} for all simulations. We assume Ca\textsuperscript{2+} concentrations prior to time t\textsubscript{0} are constant and low at \num{0.1}\unit{\micro \Molar}. All results were gathered using Matlab (MathWorks, Natick, MA).

\section*{Results}
We replace the ODEs calculating the gating variables in the six-state Siekmann model with integrodifferential equations. Each integral can be interpreted as a weighted average of \ca\ concentrations in the past. Directly translating the differential equations for the gating variables to integrals would require integrating over an interval $(-\infty, t]$ i.e. from $-\infty$ to the current time $t$. This would imply the \ipr\ has to ``average'' \ca\ concentrations spanning infinitely long in the past to the current time, or in other words, the \ipr\ must have an infinite memory of \ca\ concentrations in the past. 
Instead we assume that the \ipr\ has a ``finite memory'' and averages \ca\ concentrations over a time interval $(t-\tau, t]$ reaching $\tau$ seconds in the past. We investigate the effect of different values for the length of the time interval, $\tau$. This enables us to compare \ipr\ channels with a ``short-term memory'' i.e. small values of~$\tau$ and \ipr s with a ``long-term memory`` i.e. a large value of~$\tau$.

\subsection*{The effect of $\tau$ on Ca\textsuperscript{2+} dynamics}
An important aspect of our two-state model is the length of the time interval $\tau$. This value determines how much ``memory'' the \ipr\ has. Within our analysis, a fundamental question is: do the ion channels require ``knowledge'' of past Ca\textsuperscript{2+} concentrations to function, or is ``knowledge'' of only the present Ca\textsuperscript{2+} concentrations sufficient? 

\begin{figure}
    \centering
 \includegraphics{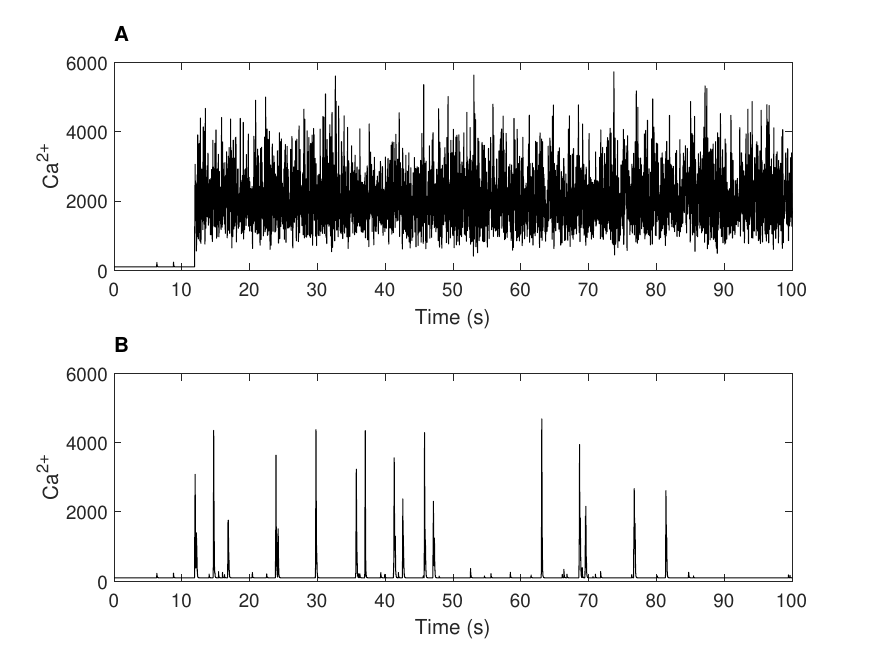}
 \caption{Comparison of Ca\textsuperscript{2+} trace for $\tau=$ \num{0.1}\unit{\second} and $\tau= $ \num{3}\unit{\second}. A: $\tau=$ \num{0.1}\unit{\second}, the Ca\textsuperscript{2+} dynamics fail with a small delay. B: $\tau=$ \num{3}\unit{\second}, Ca\textsuperscript{2+} puffs are successfully produced when $\tau$ is larger.}
    \label{Fig5}
\end{figure}

To answer this, we determined to find a threshold value for the delay length, where anything smaller than this will be detrimental to the Ca\textsuperscript{2+} dynamics. We found that when $\tau$ was set to \num{0.1}\unit{\second}, Ca\textsuperscript{2+} puffs were not produced and the ion channels stayed in a high activity mode. This suggests there is in fact a threshold below which Ca\textsuperscript{2+} puffs cannot be produced. Figure \ref{Fig5} compares the Ca\textsuperscript{2+} traces when the length of $\tau$ is \num{0.1}\unit{\second} to when it is \num{3}\unit{\second}.

\begin{figure}
    \centering
      \includegraphics[width=\linewidth]{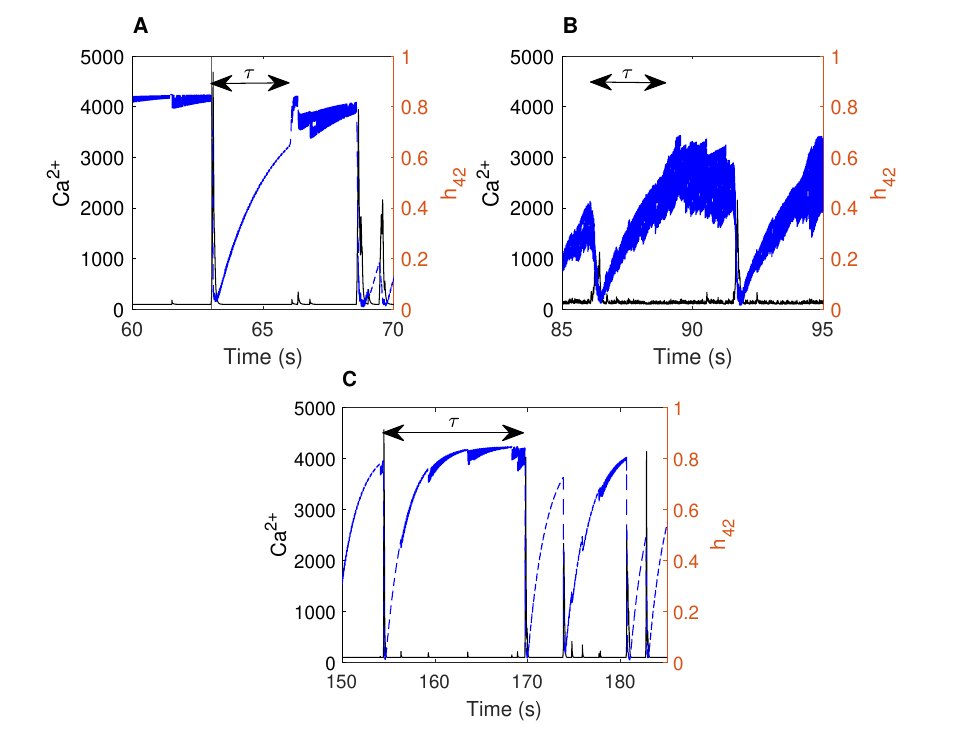}
    \caption{A: an example of the averaged $h_{42}$ gating variable and Ca\textsuperscript{2+} trace for the two-state model with $\tau=\num{3}\unit{\second}$. When the Ca\textsuperscript{2+} concentration is low, the $h_{42}$ gating variable gradually increases. When the concentration has remained constant for $\tau$ seconds, $h_{42}$ increases to equilibrium ($\sim$ \num{0.8}). It remains close to this value until a Ca\textsuperscript{2+} puff is triggered, causing the $h_{42}$ value to rapidly decrease. B: an example of the averaged $h_{42}$ gating variable and Ca\textsuperscript{2+} trace for the six-state model with $\tau=\num{3}\unit{\second}$. The $h_{42}$ value gradually increases whilst the Ca\textsuperscript{2+} concentration remains low, as seen in the two-state model. Due to the constant fluctuations in the basal Ca\textsuperscript{2+} concentration, $h_{42}$ does not reach it's equilibrium value. C: an example of the averaged $h_{42}$ gating variable and Ca\textsuperscript{2+} trace for the two-state model with $\tau=\num{15}\unit{\second}$. Increasing the length of $\tau$ produces a $h_{42}$ dynamic that is similar to that simulated by the six-state model. The $h_{42}$ gradually increases to it's equilibrium value without the sudden jump seen when $\tau$ is of a shorter length. Black full line is the Ca\textsuperscript{2+} concentration, blue dashed line is the averaged $h_{42}$ gating variable. Arrows show the length of $\tau$ following a Ca\textsuperscript{2+} puff.}
    \label{fig4}
\end{figure}

Cao et al. \cite{Cao2013} show the $h_{42}$ gating variable is important for the Ca\textsuperscript{2+} dynamics. The average $h_{42}$ gating solution for the two-state model is presented in Fig \ref{fig4}A. Whilst the Ca\textsuperscript{2+} concentration remains low at \num{0.1}\unit{\micro \Molar}, the $h_{42}$ gating variable gradually increases. If the Ca\textsuperscript{2+} concentration has remained constant for the length of $\tau$, $h_{42}$ increases to it's steady state value and remains there until a Ca\textsuperscript{2+} puff is triggered. An increase in the Ca\textsuperscript{2+} concentration causes the $h_{42}$ gating variable to decrease to a value near to zero, before gradually increasing again. We do not see this sudden increase in the $h_{42}$ gating variable for the six-state model (see Fig \ref{fig4}B) as the basal level Ca\textsuperscript{2+} concentration constantly fluctuates, therefore the $h_{42}$ gating variable never reaches equilibrium. If $\tau$ is set to a larger value, such as \num{15}\unit{\second}, the increase to equilibrium is less likely to occur, because a Ca\textsuperscript{2+} puff is usually triggered within this time frame. The $h_{42}$ dynamic resembles that of the six-state model (see Fig \ref{fig4}C). Although the $h_{42}$ dynamic changes depending on the length of $\tau$, the Ca\textsuperscript{2+} puff dynamics are not effected. It is for this reason we choose to continue our analysis using $\tau=3\unit{\second}$.

\subsection*{Introduction to the models}

We describe the steps taken to create a Ca\textsuperscript{2+} puff model based on integrodifferential equations. Puff statistics (IPI, puff amplitude and puff duration) from our model are presented. IPI distributions are parameterised using the time-dependent distribution proposed by Thurley et al. \cite{Thurley2011a} and compared. $\tau$ and $a_{\text{h}_{42}}$ are set to \num{3}\unit{\second} and \num{0.5}\unit{\second}\textsuperscript{-1}, respectively, unless stated otherwise. 



\subsubsection*{Replacing the ODEs calculating the gating variables in the Siekmann model produces equivalent results to Cao et al. \cite{Cao2013}} \label{sssec: full model}
\begin{figure}
    \centering
    \includegraphics{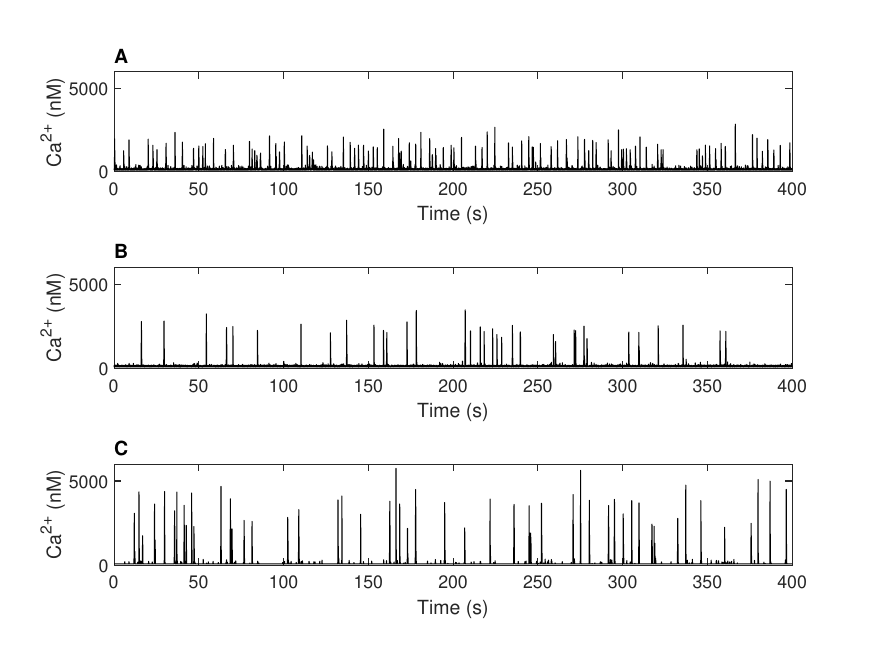}
    \caption{Examples of Ca\textsuperscript{2+} traces for models. A: Ca\textsuperscript{2+} trace produced by six-state Markov model with four integral gating variables produces equivalent results to the model by Siekmann et al. and Cao et al. \cite{Siekmann2012,Cao2013}. B: Ca\textsuperscript{2+} trace produced by a six-state Markov model with one integral gating variable. The frequency of Ca\textsuperscript{2+} puffs is reduced by the reduction of the model. C: Ca\textsuperscript{2+} trace produced by a two-state Markov model with one integral gating variable}
    \label{Fig2}
\end{figure}

Directly replacing the ODEs calculating the gating variables in the six-state Siekmann model with integrodifferential equations (Fig \ref{Fig2}A) reproduces previous results (see Cao et al. \cite{Cao2013}: Fig 2). The fitting of simulated IPI distributions to Eq \ref{eq:time dependent distribution} produced parameter values ($\lambda=0.2486$ and $\xi=0.6267$) that are similar to those described by Cao et al. \cite{Cao2013} (see \nameref{S2_Fig} for plots). Puff amplitude and duration distributions were also similar to the results produced by Cao et al. \cite{Cao2013}. 

\subsubsection*{Using quasi-steady-state approximation reduces the model whilst maintaining the correct puff dynamics}
Quasi-steady-state approximation can be used to reduce the number of gating variables in the six-state model by setting 
m\textsubscript{24}, h\textsubscript{24} and m\textsubscript{42} to their steady state. This results in a model we refer to as the ``reduced six-state model'' which comprises of six ODEs and an integral, calculating the h\textsubscript{42} gating variable. Fig \ref{Fig2}B shows an example of a Ca\textsuperscript{2+} trace produced by the reduced six-state model. In contrast to the six-state model (Fig \ref{Fig2}A), the reduced six-state model has fewer puff events, higher puff amplitudes and shorter puff duration's. Fitting of the time-dependent distribution by Thurley et al. \cite{Thurley2011a} produced parameter values of $\lambda=0.0986$ and $\xi=0.1723$, which show the average time between Ca\textsuperscript{2+} puffs is greater for the reduced mode, but puff recovery time is slower.

Cao et al. \cite{Cao2014} demonstrated that the six-state model can be reduced to a two-state model using quasi-steady-state approximation and ignoring low dwell times. We apply these methods and refer to our final model as the ``reduced two-state model''. Our results, presented in Figure \ref{Fig2}C, show that the Ca\textsuperscript{2+} traces simulated by the reduced two-state model are similar to those produced by more complex six-state models. The reduced two-state model does not have a fast lived open state---the equivalent to state $O_5$ in the six-state model---therefore the model is not able to produce openings of a small number of IP\textsubscript{3}R. This difference causes there to be less basal fluctuations in the reduced two-state model. We fit the simulated IPI distribution to the time-dependent probability density function and calculate $\lambda=0.13$ and $\xi=0.3099$. This illustrates that the frequency of Ca\textsuperscript{2+} puffs and puff recovery rate is lower for the reduced two-state model, in comparison to the six-state model.

\begin{figure}
    \centering
    \includegraphics[width=\linewidth]{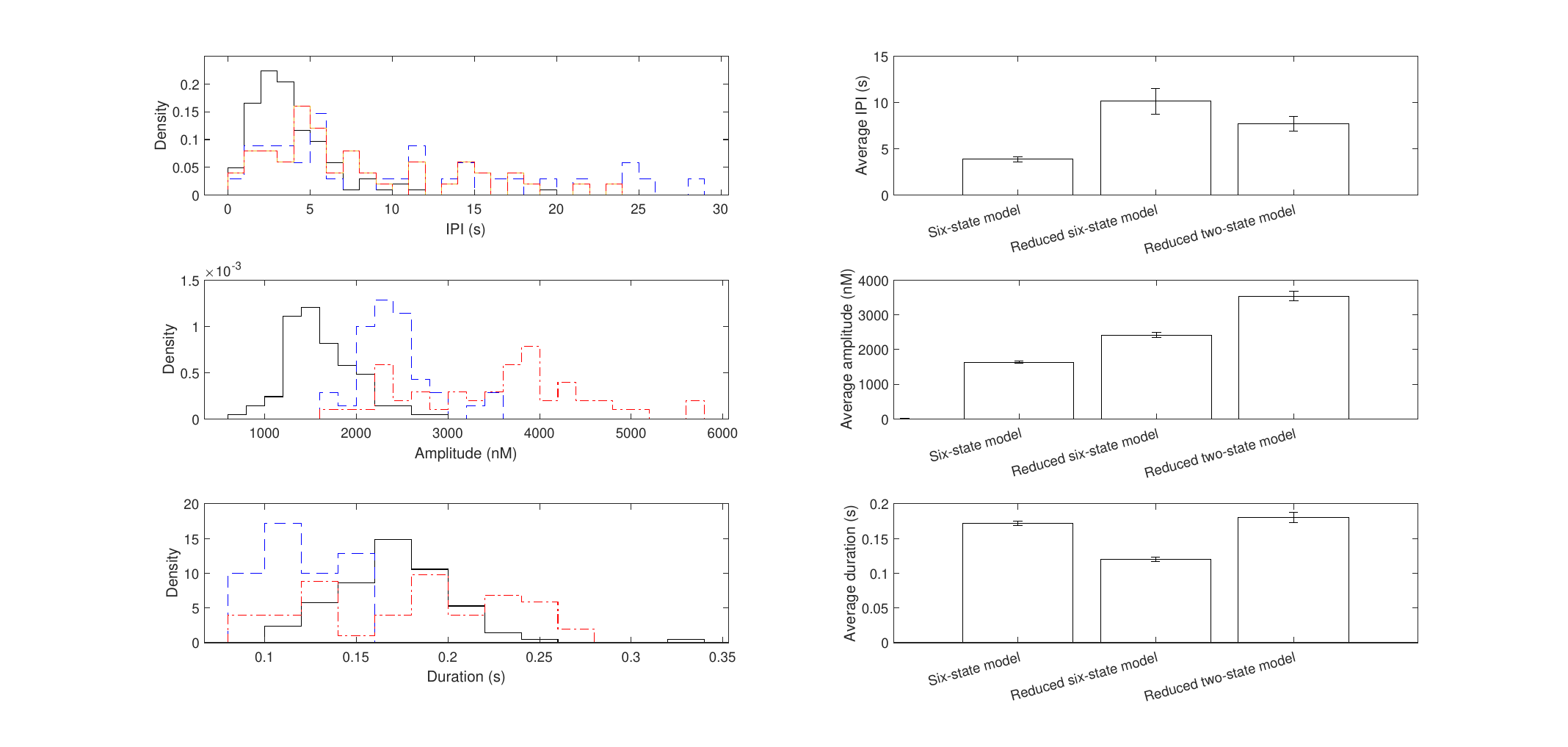}
    \caption{Comparison of average puff statistics across all three models. Bars depict the mean of each statistic $\pm$ standard error. Simplifying the six-state model using quasi-steady-state approximation leads to a decrease in the frequency of Ca\textsuperscript{2+} puff events. The increase in puff amplitude for these models implies that due to quasi-steady-state approximation a higher number of channels open at the same time, however, the channel requires a longer time period to recover from the high Ca\textsuperscript{2+} concentration and reopen.}
    \label{Fig3}
\end{figure}

Puff statistics for all models described are presented in Fig \ref{Fig3} as probability distributions and averages with standard error. The probability distributions demonstrate all models produce similar puff statistics. Reducing the model to a two-state model with one gating variable increases the average IPI and puff amplitude, however the average puff duration remains similar. Comparison of the puff statistics and averages demonstrates that the reduced two-state model can produce Ca\textsuperscript{2+} dynamics that are a good reflection of more complex models. 

\section*{Discussion}
Mathematical models simulating the Ca\textsuperscript{2+} signalling system are often complex and require a large number of parameters and equations. 

The aim of our research was to build a model for the \ipr\ that accounts for the delayed response of the channel to changes in \ca\ concentrations observed by \cite{Mak2007}. Our model is based on the hypothesis that, rather than responding only to the current \ca\ concentration~$c(t)$, the \ipr\ dynamics depends on the average~$\bar{c}(t)$ of \ca\ concentrations reaching~$\tau$ units of time in the past. Starting from the Siekmann model \cite{Siekmann2012}  
which has been shown to be incapable of generating realistic puffs if coupled directly to the time-dependent \ca\ concentration~$c(t)$ \cite{Cao2013}, we demonstrated that we can enable the model to produce puffs by replacing the dependency on~$c(t)$ by the average concentration~$\bar{c}(t)$--- provided that the length~$\tau$ of the time interval used for calculating the average \ca\ concentration~$\bar{c}(t)$ is sufficiently long. When $\tau$ was set to a small value of \num{0.1}\unit{\second} the model failed to generate Ca\textsuperscript{2+} puffs, whereas setting $\tau=\num{3}\unit{\second}$ is sufficient for enabling the model to produce puffs for the parameters chosen in Table~\ref{table: parameters}. 

This shows that a data-driven ion channel model that accounts for the delayed response to changes in ligand concentration can be constructed by first constructing a ligand-dependent infinitesimal generator~$Q(c)$ from single-channel data set at various ligand concentrations~$c$. The delayed response to changes in ligand concentrations can be incorporated into the model in a second step by parametrising the weighted average~$\bar{c}(t)$, for example, from a data set that shows rapid changes in ligand concentrations by Mak et al.\cite{Mak2007}. Thus, both data sources can be incorporated in the model separately in a transparent way. 

For the Siekmann model, a suitable choice for the weighted average~$\bar{c}(t)$ has been found by relating integral terms of the form~$\bar{c}(t)$, see~\eqref{eq:avgCa}, to the gating variables introduced to account for the delayed response to changes in \ca\ by Cao et al. \cite{Cao2013}. Following an idea demonstrated by Brady \cite{Brady1972} for the Hodgkin-Huxley model \cite{Hodgkin1952} we replaced the differential equations for the gating variables by integral terms appearing in the infinitesimal generator~$Q(\bar{c}(t)$.  
Our model successfully produced results that were comparable with those published by Cao et al. \cite{Cao2013} which was expected for infinite delay~$\tau=\infty$ because in this case, similar to the transformation of the Hodgkin-Huxley model \cite{Hodgkin1952} presented by Brady \cite{Brady1972} the models are mathematically equivalent. We then investigated how much the delay could be reduced so that the model would still produce realistic puffs. Next, we simplified our
model by using quasi-steady-state approximation to reduce the number of gating variables 
from four to one. This approach has been used 
previously, for example, see Cao et al.\cite{Cao2014} and Dupont et al. \cite{Dupont2016} and is made possible due to the rate the gating variables $m_{24}$, $h_{24}$ and $m_{42}$ reach equilibrium being so quick. The reduction in our model led to longer IPIs, higher puff amplitudes and shorter puff durations.

Finally, we followed the steps described by Cao et al. and Siekmann et al. \cite{Cao2014,Siekmann2019} to simplify our model further, reducing it to a two-state model. Our results were comparable with both the reduced six-state model and the results produced by Cao et al. \cite{Cao2014}. Such results included longer IPIs and higher puff amplitudes. Siekmann et al. \cite{Siekmann2019} state that it is not the intramodal structure of the Markov model that determines the behaviour of the ion channel, but the time-dependence of the intramode transitions. This has been shown to be true for the six and two-state models by Siekmann et al and Cao et al. \cite{Siekmann2012,Cao2013,Cao2014} and is also true for our models. We show that the behaviour and puff statistics between the six and two-state models based on integrodifferential equations are similar. However, one may argue the six-state models are a better representation of the activity within the cell because it simulates the frequent small fluctuations in Ca\textsuperscript{2+} concentration, which we do not see in the reduced two-state model.

By construction, our \ipr\ model is based on the assumption that ion channels require information of past Ca\textsuperscript{2+} concentrations. The idea that ion channels have ``memory'' of past ligand concentrations is still somewhat uncommon, for example, Villalba Galea and Chiem \cite{Galea2020} state that the activity of the ligand-gated receptor depends only on the current concentration of the agonist ligand. Interestingly, Villalba Galea and Chiem make this statement in an article where they review the evidence for memory effects in voltage-gated ion channels. However, the experiments by Mak et al. \cite{Mak2007} clearly show that the dynamics of the \ipr\ not only depends on the current concentrations of its ligands \ca\ and \ipthree\ but also on the concentrations of \ca\ and \ipthree\ that the channel has been exposed to in the past. 


We would like to consider two possible explanations for the memory effect found in the data by Mak et al. \cite{Mak2007} and represented in the architecture of our model of the \ipr. One interpretation is that the memory of the \ipr\ might have emerged due to physiological necessity---the \ipr\ is only capable of responding appropriately to variations in \ca\ concentrations if the channel ``observes'' \ca\ over the recent past. This view is supported by the dynamics of~$h_{42}$, see Fig.~\ref{fig4}. As long as no major increase in the \ca\ concentration occurs, the gating variables~$h_{42}$ of all \ipr s in the cluster continuously increase which makes the cluster of \ipr s increasingly excitable---once~$h_{42}$ has increased above a certain level, a small increase in the \ca\ concentration causes a large proportion of channels to open and release \ca, triggering a puff. In response, the gate~$h_{42}$ nearly instantaneously decreases to a value close to zero but starts to gradually increase again after the puff terminates and the \ca\ concentration has returned to the resting level. 

An alternative explanation for the memory effect is based on the biophysical basis of ``sensing'' the \ca\ concentration in the environment of the channel. Rather than being able to directly ``measure'' the \ca\ concentration, a ligand-gated ion channel like the \ipr\ has to infer the ligand concentration in its environment from the interactions of the ligand with its binding sites. Thus, rather than responding to the current \ca\ concentration $c(t)$ it is more reasonable to assume a model where the channel kinetics depends on an average \ca\ concentration~$\bar{c}(t)$ which can be related to the average time that \ca\ has been bound to the various binding sites of the channel over the course of a time interval~$\tau$.

The analysis of stochastic models is often a considerable computational challenge. Understanding stochastic dynamics generally requires time-consuming simulations which then need to be analysed statistically. However, for hybrid stochastic systems, time-dependent probability densities for the states of the Markov model depending on the continuous variables of the ODE system coupled to the Markov model can be calculated from a system of deterministic partial differential equations (PDEs). For the model presented here, we will obtain open and closed time distributions depending on \ca\ and the fluorescent buffer. Similar to the open and closed time distributions for models of single ion channels, the open and closed time densities are very useful for gaining general insights into the model behaviour. 
An interesting question to consider is how the probability density functions differ depending on how healthy the ion channel is, as demonstrated by Tveito and Lines \cite{Tveito2016}. This approach has already been applied in both cellular biology and predator prey models \cite{Tveito2016, Bressloff2018, Haw:24a}.

\section*{Supporting information}
\label{S1_Appendix}
\setcounter{figure}{0}
\renewcommand{\thefigure}{S\arabic{figure}}
\paragraph*{S1 Fig.}
\begin{figure}[h!]
    \centering
    \includegraphics[width=0.9\linewidth]{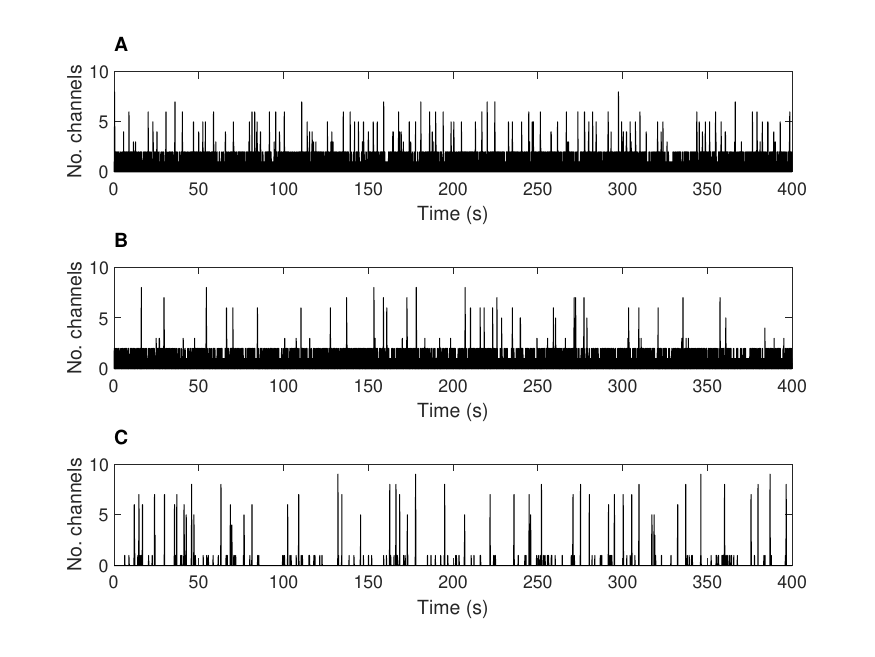}
    \caption{Examples of the number of open IP\textsubscript{3}R for each model. A: six-state model with four gating variables. B: reduced six-state model. C: reduced two-state model}
   \label{S1_Fig}
\end{figure}
\nameref{S1_Fig} shows the number of open channels for the six-state model, the reduced six-state model and the reduced two-state model using simulated Ca\textsuperscript{2+} trace data. The frequency of channels opening is higher for the six-state model (\nameref{S1_Fig}A) in comparison to the reduced six-state model (\nameref{S1_Fig}B) and the reduced two-state model (\nameref{S1_Fig}C). The lack of fast lived open state in the reduced two-state model leads to fewer fluctuations in the basal Ca\textsuperscript{2+} concentration. This is evidenced by there being less blips in \nameref{S1_Fig}C.

\paragraph*{S2 Fig.}
\label{S2_Fig}
\begin{figure}[h!]
   \centering
   \includegraphics[width=0.9\linewidth]{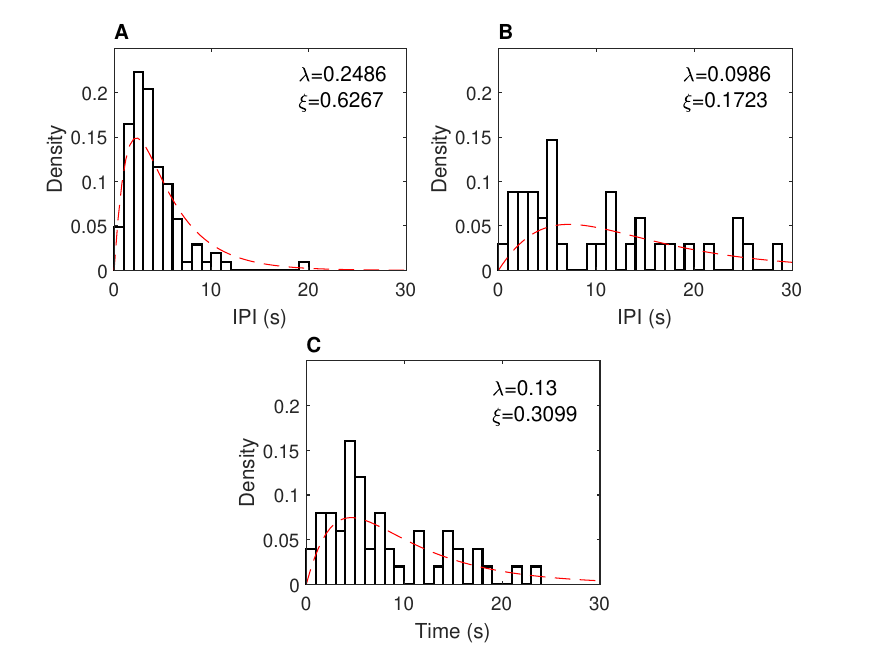}
   \caption{IPI distributions for each model fitted with the time-dependent distribution, see Eq. \eqref{eq:time dependent distribution}. A: six-state model with four gating variables. B: reduced six-state model. C: reduced two-state model}
\end{figure}

\nameref{S2_Fig} shows the IPI distributions for each model fitted with the time-dependent distribution, see Eq. \eqref{eq:time dependent distribution}.

\paragraph*{S3 Fig.}
\label{S3_Fig}
\begin{figure}[h!]
    \centering
    \includegraphics[width=0.9\linewidth]{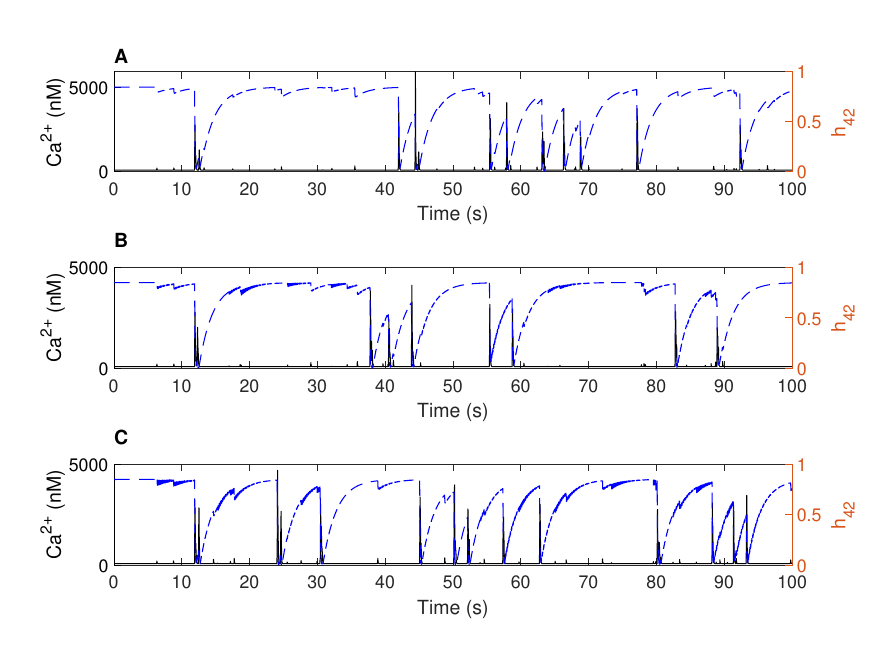}
    \caption{Choosing the most appropriate time step to calculate the $h_{42}$ integral term. A: using a time-step of $10^{-3}$\unit{\second}. B: using a time-step of $5\times10^{-3}$\unit{\second}. C: using a time-step of $10^{-2}$\unit{\second}}
\end{figure}
In the main text we showed, for the two-state model, if the length of $\tau$ was increased to \num{15}\unit{\second}, Ca\textsuperscript{2+} puff and the $h_{42}$ gating variable resembled results from the six-state model. However, computational time for simulating the integral can be expensive if $\tau$ is long. A method for reducing computational time, whilst maintaining the Ca\textsuperscript{2+} and $h_{42}$ dynamics, is to increase the time-step used to calculate the $h_{42}$ gating variable integral.

Here, we compare the average $h_{42}$ dynamic when the integral is calculated using different length time-steps i.e. using time-steps of $10^{-3}$\unit{\second}, $5\times10^{-3}$\unit{\second} and $10^{-2}$\unit{\second} with the aim of determining the maximum time-step we can use to calculate $h_{42}$ whilst maintaining the correct dynamics. We set $\tau$ to \num{15}$\unit{\second}$ as this length integral was shown to produce the desired the h\textsubscript{42} behaviour i.e. without having a sudden increase to steady state. \nameref{S3_Fig} shows that using a time-step of $10^{-2}$\unit{\second} produces similar results to when a smaller time-step is used.


\nolinenumbers

%
%
%

\bibliography{bibliography}

\end{document}